\documentclass[pra,twocolumn,aps,superscriptaddress,showpacs]{revtex4-1}

\usepackage{hyperref}
\usepackage{graphicx}
\usepackage{amsmath}
\usepackage{amsfonts}
\usepackage{amssymb}
\usepackage{epsfig}
\usepackage{subfigure}
\usepackage[usenames,dvipsnames]{color}
\usepackage{setspace}
\usepackage{bm}
\usepackage{times}
\hypersetup{
      colorlinks=true,
      citecolor=blue,
      linkcolor=blue,
      urlcolor=blue}

\begin{document}
\title{Optical lattice influenced geometry of quasi-2D binary condensates
       and quasiparticle spectra
       }

\author{K. Suthar}
\affiliation{Physical Research Laboratory,
             Navrangpura, Ahmedabad-380009, Gujarat,
             India}
\affiliation{Indian Institute of Technology,
             Gandhinagar, Ahmedabad-382424, Gujarat, India}

\author{D. Angom}
\affiliation{Physical Research Laboratory,
             Navrangpura, Ahmedabad-380009, Gujarat,
             India}

\date{\today}


\begin{abstract}

We explore the collective excitations of optical lattices filled with 
two-species Bose-Einstein condensates (TBECs). We use a set of 
coupled discrete nonlinear Schr\"odinger equations to describe the system,
and employ Hartree-Fock-Bogoliubov (HFB) theory with the Popov approximation
to analyze the quasiparticle spectra at zero temperature. The ground
state geometry, evolution of quasiparticle energies, structure of 
quasiparticle amplitudes, and dispersion relations are examined in detail.
The trends observed are in stark contrast to the case of TBECs only with
a harmonic confining potential. One key observation is the quasiparticle 
energies are softened as the system is tuned towards phase separation, but
harden after phase separation and mode degeneracies are lifted.

\end{abstract}



\maketitle

\section{Introduction}

 The experimental realization of ultracold atoms in optical lattices 
has opened up a plethora of new possibilities to study interacting quantum 
many-body systems. The optical lattices, filled with 
bosons~\cite{greiner_01,jaksh_98} or fermions~\cite{kohl_05,hofstetter_02} 
provide unprecedented precision, tunability of interactions, possibility to 
generate different geometries and mimic the external gauge fields to study
many-body systems~\cite{dalibard_11}. These are near ideal systems to observe 
quantum phenomena such as superfluidity~\cite{anderson_98,fisher_89}, 
quantum phase transition~\cite{greiner_02,altman_05}, 
Bloch oscillations~\cite{dahan_96,sorenson_98}, 
Landau-Zener tunneling~\cite{biao_00,jie_02}, and various kind of 
instabilities~\cite{konotop_02,chen_10}. In fact, the energy of collective 
excitations has emerged as fundamental and versatile tool to investigate 
many-body physics. An example of synergy between theory and experiment in
this field is the study of the effect of tunneling and mean-field interaction 
of trapped 2D optical lattices on the collective excitation. Theoretically,
Kr\"amer et al.~\cite{kramer_02} studied it in detail, and  
Fort et al.~\cite{fort_03} verified the theoretical findings in experiments. 
A detailed understanding of the excitations of superfluid phase in optical
lattices is possible with controlled variation of the lattice potential, and 
are excellent proxies to probe the properties of more complex 
condensed-matter counterparts. In this work, we examine the quasiparticle
spectrum of condensates with tight binding approximation, and the 
condensate density is described through a set of coupled discrete 
nonlinear Schr\"{o}dinger equations.

 The introduction of a second species in the optical lattices,
two-species BECs (TBECs) in lattices, creates a versatile model to probe 
diverse phenomena in physics. These are promising candidates to explain 
phenomena associated with fermionic correlations~\cite{paredes_03}, phase 
separation~\cite{alon_06}, hydrodynamical instability~\cite{lundh_12} and 
novel phases~\cite{duan_03,kuklov_03}. One remarkable property of 
TBECs is the phase segregation, which occurs when the interspecies interaction 
is stronger than the geometric mean of the intraspecies 
interactions~\cite{lun_96}. To date, TBECs in optical lattices have been 
experimentally realized in two different atomic species~\cite{catani_08} and 
two different hyperfine states of the same atomic 
species~\cite{gadway_10,soltan_11}. 
It must be emphasized that TBECs with harmonic potential only have been
realized in two different species of
alkali-atoms~\cite{modugno_02,lercher_11,mccarron_11,thalhammer_08}, and in 
two different isotopes~\cite{papp_08}, and in two different hyperfine 
states~\cite{myatt_97,stamper_98,hall_98,tojo_10}.
These experiments have examined phase 
separation and other phenomena which are unique to binary BECs.
The phenomenon of phase separation and transition from miscible-to-immiscible 
or vice versa has also been the subject of several 
theoretical studies~\cite{esry_97,ohberg_98,svidzinsky_03,arko_14a}.
These recent developments are motivations to probe the rich physics 
associated with TBECs in optical lattices. In recent works, we  
have investigated the fluctuation induced instability of dark solitons in 
TBECs~\cite{arko_14b} and change in the topology of the TBECs in 
quasi-1D lattices~\cite{suthar_15}. However, to study the effects of 
fluctuations, either quantum or thermal, in optical lattices filled with
TBECs it is essential to have a comprehensive understanding of the 
quasiparticle spectra.

In this paper, we examine the evolution of the quasiparticle spectra
of TBECs in quasi-2D optical lattices at zero temperature. For this we use 
HFB formalism with Popov approximation, and tune one of the inter-atomic
interactions to drive the TBEC from miscible to immiscible phase. In 
the immiscible domain, we show that the ground state has {\em side-by-side} 
density profile. This is in contrast to the case of quasi-1D system, where
the ground state has {\em sandwich} density profile. To identify the 
geometry of the ground state, we examine the quasiparticle spectra using
Bogoliubov de Gennes (BdG) analysis. For a stable ground state configuration,
the spectra is real, but complex for metastable states. Following BdG 
analysis, we further examine the dispersion relation of binary system 
in optical lattices. These relations are used to understand the 
structure of the lower and higher energy excitations for miscible 
and immiscible domain of TBEC in lattice system.
The dispersion relations are important to understand the nature of the
excitations~\cite{kurn_99,steinhauer_02,ticknor_14}, and Bragg 
spectroscopy~\cite{du_10} of ultracold quantum gases. These spectroscopic 
studies present full momentum-resolved measurements of the band structure and 
the associated interaction effects at several lattice depths~\cite{ernst_10}.
In fact, these relations have proved the presence of the rotonlike excitation
in trapped dipolar BECs~\cite{wilson_10,ticknor_11,bisset_13,blakie_13}.

The paper is organized as follows: In Sec.~\ref{theory_2s2d}, we describe
the HFB-Popov formalism and the dispersion relations for TBEC confined in 
optical lattices. The quasiparticle mode evolution and characteristic of the 
quasiparticle excitations with dispersion curves are presented in
Sec.~\ref{results}. Finally, we conclude with the key finding of the present 
work in the Sec.~\ref{conc}.


\section{Theory and methods}
\label{theory_2s2d}

Consider TBEC of dilute atomic gases in an optical lattice with 
a harmonic oscillator potential as a confining envelope potential. So, 
the net external potential is
\begin{eqnarray}
 V^{k} (\mathbf r)  &=& V^{k}_{\rm ho} + V^{k}_{\rm latt} \nonumber \\
    &=& \frac{m_k}{2} (\omega_{x}^2  x^2 + \omega_{y}^2 y^2 
           + \omega_{z}^2 z^2) + V_0 [\sin^2(2\pi x/\lambda_L) \nonumber \\
           &&+ \sin^2(2\pi y/\lambda_L) + \sin^2(2\pi z/\lambda_L)],
\end{eqnarray}
where $k = 1,2$ denotes the species index, $m_k$ is the atomic mass of 
the $k$th species, $\omega_i(i=x,y,z)$ are the frequencies of the harmonic 
potential along each direction, $V_0 = sE_R$ is the depth of the lattice 
potential in terms of the recoil energy $E_R = \hbar^2 k^2_{L}/2m$ and 
dimensionless scale factor $s$. Here, $k_L = 2\pi/\lambda_L$ is the wave 
number of the laser beam with wavelength $\lambda_L$ used to generate the
optical lattice, and hence  the lattice constant of the system is
$a = \lambda_L/2$. It is to be noted that we consider the same external 
potential for both the condensate, and at $T = 0$ K the grand canonical 
Hamiltonian of the system is
\begin{eqnarray}
\hat{H}& = &\sum_{k=1}^{2} \int d\mathbf r~\hat{\Psi}^{\dagger}_{k}(\mathbf r)
           \bigg [-\frac{\hbar^2 {\nabla}^2}{2 m_k} + V^{k}(\mathbf r) - \mu_k
         + \frac{U_{kk}}{2}\hat{\Psi}^{\dagger}_{k}(\mathbf r) \nonumber \\
           &\times& \hat{\Psi}_{k}(\mathbf r) \bigg ]\hat{\Psi}_{k}(\mathbf r)
            + U_{12} \int d\mathbf r~\hat{\Psi}^{\dagger}_{1}(\mathbf r)
            \hat{\Psi}^{\dagger}_{2}(\mathbf r)
            \hat{\Psi}_{1}(\mathbf r)\hat{\Psi}_{2}(\mathbf r),
            \nonumber\\
\label{grand_can}
\end{eqnarray}
where $\hat{\Psi}_{k}$, $\mu_k$ and $U_{kk}$ are the bosonic field operator,
chemical potential and intraspecies interaction strength of 
$k$th species, and $U_{12}$ is the interspecies interaction strength. 
In the present study, we consider all the interactions to be repulsive, that is
$U_{kk},U_{12}>0$. If the lattice is deep, i.e. $V_0\gg \mu_k$, the 
tight binding approximation (TBA) is applicable, and bosons occupy only the 
lowest energy band. In this approximation, the condensate is well localized 
within each lattice site, and the field operator for each of the 
species can be written as
\begin{equation}
 \hat{\Psi}_{k}(\mathbf r) = \sum_{\xi} \hat{a}_{k\xi} \phi_{k\xi}(\mathbf r),
\label{tba}
\end{equation} 
where $\hat{a}_{k\xi}$ is the annihilation operator of the $k$th species 
at the lattice site with identification index $\xi$, which is a unique 
combination of the lattice index along $x$, $y$ and $z$ axes.
The basic element of TBA lies in the definition of $\phi_{k\xi}(\mathbf r)$,
these are orthonormalized on-site Gaussian wave functions localized 
at the $\xi$th lattice site. Using the above definition of 
$\hat{\Psi}_{k}(\mathbf r)$ in Eq.~(\ref{grand_can}), we get the Bose-Hubbard 
Hamiltonian (BH) of the system.


\subsection{HFB-Popov approximation for quasi-2D TBEC in optical lattices}

 To create a potential suitable to generate quasi-2D TBEC in optical lattices,
set the frequencies to satisfy the condition 
$\omega_x = \omega_y = \omega_{\perp} \ll \omega_z$. The excitations along the
tight or high frequency, $z$-axis, are of higher energies and we consider
the condensate is in ground state along the $z$-axis at low
temperatures $T\ll \hbar\omega_z/k_B$ with $k_B$ as the Boltzmann constant. 
Hence, the excitations of importance for quantum and thermal 
fluctuations are along the radial direction. In the TBA, 
the BH Hamiltonian which describes the system is 
\begin{eqnarray}
\hat{H} = && \sum_{k=1}^2 \left[- J_k \sum_{\langle \xi\xi'\rangle} 
             \hat{a}^{\dagger}_{k\xi}\hat{a}_{k\xi'} 
           + \sum_\xi(\epsilon^{(k)}_{\xi} - \mu_k)
              \hat{a}^{\dagger}_{k\xi}\hat{a}_{k\xi}\right] \nonumber\\
          &+& \frac{1}{2}\sum_{k=1}^{2} U_{kk}\sum_\xi \hat{a}^{\dagger}_{k\xi}
              \hat{a}^{\dagger}_{k\xi}\hat{a}_{k\xi}\hat{a}_{k\xi} \nonumber\\
          &+& U_{12} \sum_\xi \hat{a}^{\dagger}_{1\xi}\hat{a}_{1\xi}
              \hat{a}^{\dagger}_{2\xi}\hat{a}_{2\xi},
\label{bh2d}              
\end{eqnarray}
where the index $\xi$ covers all the lattice sites. The summation index 
$\langle \xi\xi'\rangle$ represents the nearest-neighbour, for illustration
take $\xi\equiv(i,j)$ with $i$ and $j$ as labels of a lattice site along $x$ 
and $y$ axes, respectively. The possible values of $\xi'$ in 
$\langle \xi\xi'\rangle$ are then $(i-1, j)$, $(i+1, j)$, $(i, j-1)$, and 
$(i, j+1)$. The operator $\hat{a}_{k\xi}(\hat{a}^{\dagger}_{k\xi})$ is the 
bosonic annihilation (creation) operator of the $k$th species at the $\xi$th 
lattice site, and $J_k$s are the tunneling matrix elements. The effect of the 
envelope harmonic trapping potentials is subsumed in the offset energy 
$\epsilon^{(k)}_{\xi} = \Omega(i^2 + j^2)$. Here,
$\Omega = m\omega^2_{\perp}a^2/2$ is the strength of the harmonic 
confinement. For simplicity, we assume the 
tunneling strength of the two species are identical in both $x$ and $y$ axes.
For large tunneling strength and density, $J_k \gg \nu U_{kk}, \nu U_{12}$ 
with $\nu$ as the filling factor, the bosons remain in superfluid phase. In 
this regime, the equilibrium properties of the system at $T = 0$ K is well 
described by the 2D coupled discrete nonlinear Schr\"odinger equations (DNLSEs)
\begin{subequations}
 \begin{eqnarray}
  \mu_1 c_\xi = &-& J_1 \sum_{\xi'} c_{\xi'} + \left [ \epsilon^{(1)}_\xi 
               + U_{11} n^{c}_{1\xi} + U_{12} n^{c}_{2\xi} \right ]  c_\xi,
                      \nonumber \\~\\
  \mu_2 d_\xi = &-& J_2 \sum_{\xi'} d_{\xi'} + \left [ \epsilon^{(2)}_\xi 
               + U_{22} n^{c}_{2\xi} + U_{12} n^{c}_{1\xi} \right ] d_\xi, 
                      \nonumber \\
 \end{eqnarray}
 \label{dnls2d}
\end{subequations}
\noindent where $c_{\xi}\equiv c_{i,j}$ and $d_{\xi}\equiv d_{i,j}$ are the 
complex amplitudes associated with the condensate wave functions of each 
species, and satisfy the normalization conditions 
$\sum_\xi|c_\xi|^2= \sum_\xi|d_\xi|^2=1$. The summation $\xi'$ is over the 
nearest neighbours to the site $\xi$, more explicitly
\begin{equation}
\sum_{\xi'} c_{\xi'} \equiv c_{\xi-1} + c_{\xi+1} 
                  \equiv c_{i-1,j} + c_{i+1,j} + c_{i,j-1} + c_{i,j+1}.
\end{equation}
 From the definition of $\phi_{k\xi}$, in Eq.(\ref{dnls2d}) 
$n^{c}_{1\xi} = |c_\xi|^2$ and  $n^{c}_{2\xi} = |d_\xi|^2$ are 
the condensate densities of the first and second species at the $\xi$th
lattice site, respectively. In the Bogoliubov approximation, we define the 
annihilation operators as
$\hat{a}_{1\xi} = (c_{\xi} +
\hat{\varphi}_{1\xi})e^{-i \mu_1 t/\hbar}$, $\hat{a}_{2\xi} = (d_{\xi} +
\hat{\varphi}_{2\xi})e^{-i \mu_2 t/\hbar}$, and the new definition of the
creation operators are the hermitian conjugates. The operator parts,  
($\hat{\varphi}_{1\xi}$ or $\hat{\varphi}_{2\xi})$
represent small perturbations, and identify with the quantum and thermal
fluctuations in the system. This approximation, when used in Eq.~(\ref{bh2d}), 
partition the BH Hamiltonian to terms of different orders in the fluctuation
operators. The lowest (zeroth) order term leads to the time-independent 
DNLSEs [Eq.~(\ref{dnls2d})]. The leading order correction terms, linear
in $\hat{\varphi}$, describe the effects arising from quantum and thermal
fluctuations of the system. A more detailed description of the derivation 
is given in one of our previous works~\cite{suthar_15}. The normal modes of 
the fluctuations, or the quasiparticle operators are defined through the 
Bogoliubov transformation 
\begin{subequations}
 \begin{eqnarray}
  \hat\varphi_{k\xi} &=& \sum_l\left[u^l_{k\xi}\hat{\alpha}_l e^{-i \omega_l t} 
             - v^{*l}_{k\xi}\hat{\alpha}^{\dagger}_l e^{i \omega_l t}\right],\\
   \hat\varphi^{\dagger}_{k\xi} &=& \sum_l\left[u^{*l}_{k\xi}
                                  \hat{\alpha}^{\dagger}_l e^{i \omega_l t} 
                         - v^{l}_{k\xi}\hat{\alpha}_l e^{-i \omega_l t}\right],
 \label{bog_trans_2d}                        
 \end{eqnarray}
\end{subequations}
where $u^l_{k\xi}$ and  $v^l_{k\xi}$ are the quasiparticle amplitudes for the
$k$th species in quasi-2D optical lattice potential, and 
$\omega_l = E_l/\hbar$ is the frequency of the $l$th quasiparticle mode
with $E_l$ as the mode excitation energy. Further more, the quasiparticle
amplitudes satisfy the normalization condition
\begin{equation}
\sum_{k\xi}\left(u^{*l}_{k\xi} u^{l'}_{k\xi}
 - v^{*l}_{k\xi} v^{l'}_{k\xi} \right) = \delta_{ll'}.
\end{equation}
Here $\hat{\alpha}_l (\hat{\alpha}^{\dagger}_l)$ are the quasiparticle
annihilation (creation) operators, which satisfy the Bose commutation
relations. The above transformation diagonalizes the BH Hamiltonian, and
taking into account the terms of higher order in fluctuation operators
in total Hamiltonian leads to the HFB-Popov equations 
\begin{subequations}
 \begin{eqnarray}
  E_l u^l_{1,\xi} = &-& J_1(u^l_{1,\xi-1} + u^l_{1,\xi+1}) 
                        + \mathcal{U}_1 u^l_{1,\xi} 
                        - U_{11} c^2_\xi v^l_{1,\xi} \nonumber\\ 
                    &+& U_{12} c_\xi(d^{*}_\xi u^l_{2,\xi} 
                        - d_\xi v^l_{2,\xi}),\\
  E_l v^l_{1,\xi} = &~& J_1(v^l_{1,\xi-1} + v^l_{1,\xi+1}) 
                        + \underline{\mathcal{U}}_1
                        v^l_{1,\xi} + U_{11} c^{*2}_\xi u^l_{1,\xi} \nonumber\\
                    &-& U_{12} c^{*}_\xi(d_\xi v^l_{2,\xi} 
                        - d^{*}_\xi u^l_{2,\xi}),\\
  E_l u^l_{2,\xi} = &-& J_2(u^l_{2,\xi-1} + u^l_{2,\xi+1}) 
                        + \mathcal{U}_2 u^l_{2,\xi} 
                        - U_{22} d^2_\xi v^l_{2,\xi} \nonumber\\ 
                    &+& U_{12} d_\xi(c^{*}_\xi u^l_{1,\xi} 
                        - c_\xi v^l_{1,\xi}),\\
  E_l v^l_{2,\xi} = &~& J_2(v^l_{2,\xi-1} + v^l_{2,\xi+1}) 
                        + \underline{\mathcal{U}}_2
                        v^l_{2,\xi} + U_{22} d^{*2}_\xi u^l_{2,\xi} \nonumber\\
                    &-& U_{12} d^{*}_\xi(c_\xi v^l_{1,\xi} 
                        - c^{*}_\xi u^l_{1,\xi}),
 \end{eqnarray}
 \label{bdg_eq_2sp}                 
\end{subequations}
where $\mathcal{U}_1 = 2 U_{11} (n^{c}_{1\xi} + \tilde{n}_{1\xi}) 
+ U_{12} (n^{c}_{2\xi} + \tilde{n}_{2\xi}) + (\epsilon^{(1)}_\xi - \mu_1)$,
$\mathcal{U}_2 = 2 U_{22} (n^{c}_{2\xi} + \tilde{n}_{2\xi}) 
+ U_{12} (n^{c}_{1\xi} + \tilde{n}_{1\xi}) + (\epsilon^{(2)}_\xi - \mu_2)$ with
$\underline{\mathcal{U}}_k = -\mathcal{U}_k$. The density of the noncondensate
atoms at the $\xi$th lattice site is
\begin{equation}
 \tilde{n}_{k\xi} = \sum_l [ (|u^l_{k\xi}|^2 + |v^l_{k\xi}|^2)N_0(E_l) 
                   + |v^l_{k\xi}|^2],
\end{equation}
with $N_0(E_l)$ as the Bose-factor of the system with energy $E_l$
at temperature $T$. The last term in the $\tilde{n}_{k\xi}$ is quantum
fluctuations which is independent of the Bose-factor, and hence
represents the quantum fluctuations of the system.


\subsection{Dispersion relations of binary BEC}

The dispersion relations, in general, determines how a system responds to
external perturbations. So, in TBECs in optical lattices as well, it is 
important to examine the dispersion relations to understand how the system 
evolves after applying an external perturbation. Examples of 
current interest are topological defects generated through phase 
imprinting, evacuating single or multiple lattice sites, and tuning the 
lattice or harmonic potential parameters. To study the dispersion relation
of the quasiparticles in optical lattices with a background trapping 
potential, we follow the definition in Ref.~\cite{wilson_10}. Following
which, we take the Fourier transform of the quasiparticle 
amplitudes, and compute the expectation value of the linear momentum 
$\langle k_{\xi}\rangle$ of each quasiparticle. So, in the 
momentum-space representation, for the $l$th quasiparticle
\begin{equation}
 {\langle k_{\xi}\rangle}_l = \left[\frac{\sum_{\alpha} 
                                         \int d{{\bf k_\xi}} k^2_{\xi} 
     [\tilde{u}^l_{\alpha} ({\bf k_\xi}) + \tilde{v}^l_{\alpha} ({\bf k_\xi})]}
     {\sum_{\alpha} \int d{{\bf k_\xi}} [\tilde{u}^l_{\alpha} ({\bf k_\xi}) 
                         + \tilde{v}^l_{\alpha} ({\bf k_\xi})]}\right]^{1/2},
\label{wave_vec}
\end{equation}
where $k_{\xi} = (k_{i},k_{j})$ is the lattice site dependent wave-number and
$\alpha = 1,2$ is the index for species. Here 
$\tilde{u}^l_{\alpha} ({\bf k_\xi}) = \mathcal{F}[{u}^l_{\alpha}(\xi)]$, and
$\tilde{v}^l_{\alpha} ({\bf k_\xi}) = \mathcal{F}[{v}^l_{\alpha}(\xi)]$ are the 
lattice site dependent quasiparticle amplitudes in momentum space, with 
$\mathcal{F}$ representing the Fourier transform. We, then, determine the
discrete form of the dispersion relation by associating 
$\langle k_{\xi}\rangle_l$ to the excitation energies $E_l$. For TBECs
in harmonic potential the dispersion curves were examined in a previous work, 
and reported unique trends in the miscible and immiscible
regimes~\cite{ticknor_14}. Compared to which the presence of the optical 
lattice potential is expected to modify the dispersive properties of the 
systems in the present study. To examine the differences, and identify
unique trends we compute $\langle k_{\xi}\rangle_l$ and study the dispersion
curves in miscible and immiscible domains.


\subsection{Numerical methods}

 To solve the coupled DNLSEs in Eq.~(\ref{dnls2d}) at $T=0$ K, we first scale
the equations and rewrite in dimensionless form~\cite{suthar_15}. The 
equations are then solved using the fourth order Runge-Kutta method. For the
zero temperature computations we begin by neglecting the noncondensate 
density ($\tilde{n}_{k\xi}$) at each lattice site, and choose the 
initial guess values of the complex amplitudes with Gaussian or 
side-by-side envelope profile such that the quasiparticle energy spectrum is 
real. To obtain ground state of the system, we solve the DNLSEs 
with imaginary-time propagation. As described earlier, in the TBA, we take 
a basis set consisting of orthonormalized Gaussian functions
localized at each lattice site. Hence, the basis set size or the number of 
basis functions is equal to the number of lattice sites in the system. 
Furthermore, to obtain the excitation spectrum we cast HFB-Popov 
Eqs.~(\ref{bdg_eq_2sp}) as a matrix eigenvalue equation. For the computations
at $T = 0$ K, the matrix is diagonalized using the routine ZGEEV, routine 
to diagonalize non-symmetric matrix with complex elements, from the LAPACK 
library~\cite{anderson_99} to obtain the quasiparticle energies and 
amplitudes $E_l$, and $u^l_\xi$'s and $v^l_\xi$'s, respectively. However,
when $T \neq 0$ K a larger number of basis functions is required to obtain
a correct description of the thermal fluctuations, and this increases
the dimension of the matrix corresponding to Eqs.~(\ref{bdg_eq_2sp}). It
is then better to use ARPACK~\cite{arpack} library for diagonalization as it 
is faster, and provides the option to compute a limited set of eigenvalues
and eigen functions. The other advantage of using ARPACK is the optimal
storage of large sparse matrices. In the latter part of our work to 
compute the dispersion curves, which in the present approach require 
quasiparticle amplitudes in the momentum 
representation, we use the FFTW library~\cite{fftw3_05} in Intel MKL. 

\begin{figure}[ht]
 {\includegraphics[width=8.5cm] {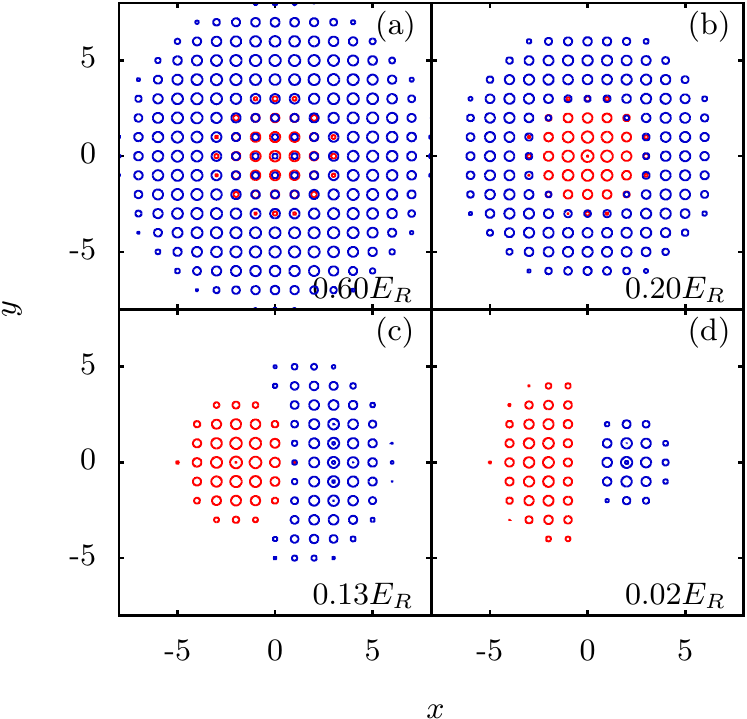}}
  \caption{The geometry of the condensate density profiles and its transition
           from miscible to the immiscible domain for $^{87}$Rb -$^{85}$Rb
           TBEC. (a) At higher $U_{22}$, the density of both species partially
           overlap, (b) as we decrease $U_{22}$ it changes into sandwich-type
           profile. At a critical value of $U_{22}$ ($0.16 E_R$), both
           condensate segregate and rotational symmetry is broken, which 
           results in side-by-side density profile in immiscible domain
           shown in (c,d). 
           Here species labeled $1(2)$ is shown as red (blue) contours.}
  \label{den_rb}
\end{figure}


\section{Results and discussions}
\label{results}

 To examine the mode evolution of quasi-2D TBEC in optical lattices, we 
consider two cases from the experimentally realized TBECs, 
$^{87}$Rb - $^{85}$Rb~\cite{papp_08} and $^{133}$Cs - $^{87}$Rb
~\cite{mccarron_11,lercher_11}. The former and latter are examples of TBECs 
with negligible, and large mass differences between the species, respectively. 
Another basic difference is, starting from miscible phase, the 
passage to the immiscible phase. In the  $^{87}$Rb - $^{85}$Rb TBEC, the 
background scattering length of $^{85}$Rb is negative, and hence to obtain 
stable $^{85}$Rb condensate~\cite{cornish_00} it is essential to render it 
repulsive using magnetic Feshbach resonance~\cite{courteille_98,roberts_98}. 
The same can be employed to drive the system from miscible to immiscible 
domain. On the other hand, in $^{133}$Cs - $^{87}$Rb TBEC, the 
inter-species scattering length is tuned through a magnetic Feshbach
resonance~\cite{pilch_09} to steer the TBEC from miscible to immiscible 
domain or vice-versa.

 For the $^{87}$Rb - $^{85}$Rb TBEC, we assume $^{87}$Rb and $^{85}$Rb as 
the first and second species, respectively. For simplicity, and ease of
comparison without affecting the results, the radial trapping frequency 
of the two species are chosen to be identical 
$\omega_x = \omega_y = 2\pi\times 50$ Hz, with $\omega_\perp/\omega_x=20.33$. 
The wavelength of the laser beam to create the 2D lattice potential and 
lattice depth are $\lambda_L = 1064$ nm and $V_0 = 5 E_R$, respectively.  To 
improve convergence, and have a good description of the optical lattice 
properties, we take the total number of atoms $N_{1} = N_{2} = 300$ confined 
in a ($30\times30$) lattice system. We use these set of parameters to study
the $^{133}$Cs - $^{87}$Rb TBEC as well.


\subsection{Mode evolution of trapped TBEC at $T=0$ K}

 To solve the DNLSE we consider Gaussian basis function of width $0.3a$, where
$a$ is the lattice constant, to evaluate the lattice parameters. In the 
case of $^{87}$Rb - $^{85}$Rb TBEC, the tunneling matrix elements are
$J_1 = 0.66 E_R$ and $J_2 = 0.71 E_R$, and $U_{11}$ = $ 0.07 E_R$ and 
$U_{12}$ = $0.15 E_R$ are the intraspecies and interspecies interactions,
respectively. The difference in the values of $J_1$ and $J_2$ arises from
the mass difference of the species in the TBEC system. Following the same 
steps, the parameters for the $^{133}$Cs - $^{87}$Rb TBEC are
$J_1 = 0.66 E_R$,  $J_2 = 1.70 E_R$, $U_{11}$ = $0.96 E_R$ and 
$U_{22}$ = $0.42 E_R$. In both the cases, we drive the system from miscible
to immiscible phase, and examine the evolution of the modes in detail.

\begin{figure}[ht]
 {\includegraphics[width=8.5cm] {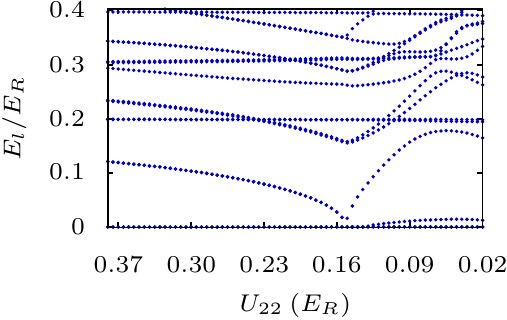}}
  \caption{The evolution of the low-lying quasiparticle modes as a function of 
           the intraspecies interaction ($U_{22}$) for the $^{87}$Rb-$^{85}$Rb 
           TBEC held in quasi-2D optical lattices. Here $U_{22}$ is in units 
           of the recoil energy $E_R$.}
  \label{mode_rb_en}
\end{figure}


\subsubsection{$^{87}$Rb - $^{85}$Rb TBEC}

As mentioned earlier $U_{22}$, the intraspecies interaction of $^{85}$Rb, is
decreased to drive the TBEC from miscible to immiscible domain. The changes 
in the ground state density profile are shown in Fig.~\ref{den_rb}. In the
miscible domain, the profiles overlap and there is a shift in the position
of the density maxima as $U_{22}$ is decreased [Fig.~\ref{den_rb}(b)].
At a critical value $U_{22}^{\rm c}$, the two species undergo phase separation
with side-by-side density profiles and breaks the rotational symmetry.
The features of the quasiparticles too change in tandem with the 
density profile, and the variation of the excitation energies with 
$U_{22}$ are shown in Fig.~\ref{mode_rb_en}. 
\begin{figure}[ht]
 {\includegraphics[width=8.5cm] {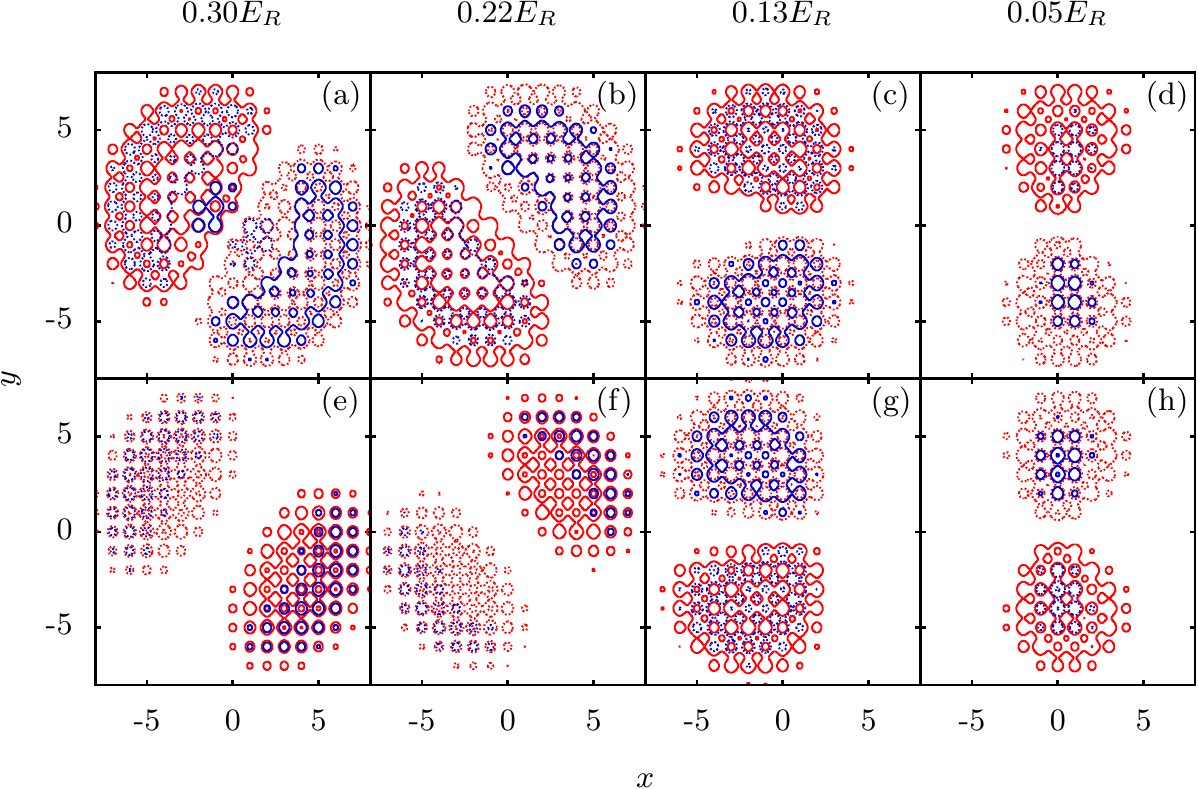}}
  \caption{The evolution of quasiparticle amplitude corresponding to the
           slosh mode of first 
           species (a-d) and second species (e-h) in $^{87}$Rb -$^{85}$Rb TBEC 
           as $U_{22}$ is decreased from $0.30 E_R$ to $0.05 E_R$. The value of 
           $U_{22}$ is shown at the top of the figures. Here the red 
           contours represent the quasiparticle amplitude ($u_{1}(x,y)$ and 
           $u_{2}(x,y)$), whereas the blue contours represent the quasihole 
           amplitude ($v_{1}(x,y)$ and $v_{2}(x,y)$). The density perturbation 
           is from dotted contours to the solid contours. 
           }
  \label{mode_fn1}
\end{figure}
\begin{figure}[ht]
 {\includegraphics[width=8.5cm] {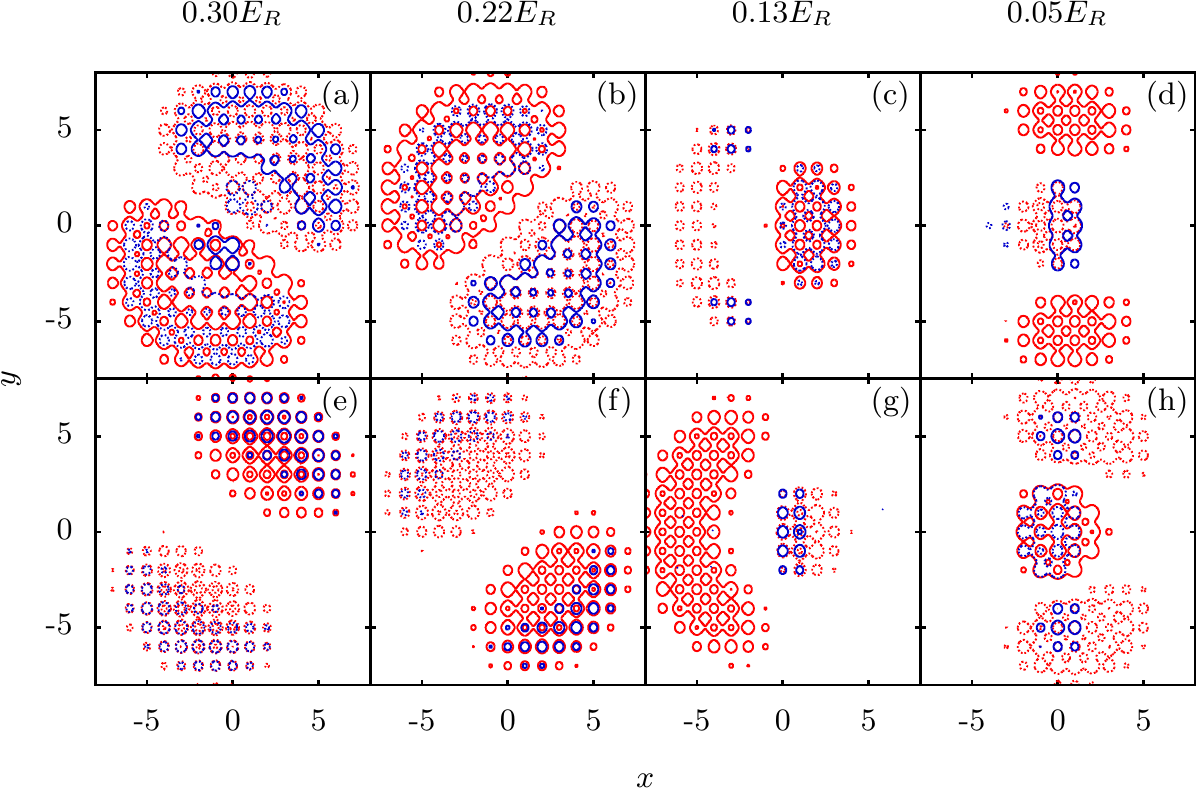}}
   \caption{The evolution of the quasiparticle amplitude corresponding to
           the other slosh mode, 
            which is degenerate to the mode shown 
            in Fig.~\ref{mode_fn1} in miscible domain. These amplitudes 
            correspond to the first species (a-d) and second species (e-h) of 
            $^{87}$Rb -$^{85}$Rb TBEC with the change in $U_{22}$, which is 
            shown at the top of the figures. At a critical value of $U_{22}$, 
            this mode hardens and gets transformed into an interface mode 
            (d, h). Here the red contours represent the quasiparticle 
            amplitude, whereas the blue contours represent the quasihole 
            amplitude.  
            }
  \label{mode_fn2}
\end{figure}
To obtain the mode evolution curves, we do a series of computations starting 
from the miscible domain of the system (higher $U_{22}$), and decrease 
$U_{22}$ to values below $U_{22}^{\rm c}$. 

In the miscible domain, all the excitation modes are doubly degenerate. 
As $U_{22}$ is lowered, eigen energies of modes with different
phases of $u_1$ and $u_2$, or out-of-phase modes
decrease in energy, and degeneracy is lifted when $U_{22}$ is 
below $U_{22}^{\rm c}$. The slosh and Kohn modes are the two lowest energy
ones in the miscible domain, and are associated with the out-of-phase 
and in-phase modes, respectively. The structure of the two degenerate
slosh modes are shown in Fig.~\ref{mode_fn1}(a,b,e,f) and 
Fig.~\ref{mode_fn2}(a,b,e,f). In general, the doubly degenerate modes are 
$\pi/2m$ rotation of each other, where $m$ is the azimuthal quantum number.
For the slosh modes this property is evident from the figures. One of the
degenerate slosh modes goes soft at $U_{22}^{\rm c} = 0.16$,
in particular, it is the one which is in-phase with the condensate 
density, but the other slosh mode gains energy at phase separation. Thus, 
below $U_{22}^{\rm c}$ the degeneracy of the slosh modes is lifted.
On further decrease of $U_{22}$ one striking effect of the optical lattice 
potential is observed: the soft slosh mode gains energy and is 
transformed into an interface mode. This is in stark contrast to the case 
without the lattice potential, where the mode remains soft~\cite{ticknor_13}. 
This is also apparent from the nature of the quasiparticle amplitudes shown 
in the figures. The Kohn mode, on the other hand, remains steady with an 
energy of $0.2 E_R$. 
 
 Considering the general trend, there are only mode crossings in the miscible 
domain, however, both mode crossing and avoided crossings occur in the 
phase-separated domain. Prior to phase separation, out-of-phase
modes decrease in energy as $U_{22}$ is lowered, but the 
in-phase modes remain steady. So, no mode mixing occurs
when modes of the former type encounters the latter, and they cross each 
other. However, when $U_{22}$ is below the critical value, degeneracies are 
lifted, and mode mixing can occur. This explains the presence of avoided 
crossings in the phase-separated domain. The energies of the out-of-phase 
modes decrease monotonically with decrease in $U_{22}$ as it favours 
phase separation. After phase separation, these modes get hardened due to 
rotational symmetry breaking. It must be noted that, as shown in Fig.  
~\ref{den_rb}(b), the density profiles are shell structured or rotationally 
symmetric for intermediate values of $U_{22}$. However, there is a sharp
transition to side-by-side density profile as phase-separation occurs
when $U_{22}$ is lowered.


\subsubsection{$^{133}$Cs - $^{87}$Rb TBEC}
\begin{figure}[ht]
 {\includegraphics[width=8.5cm] {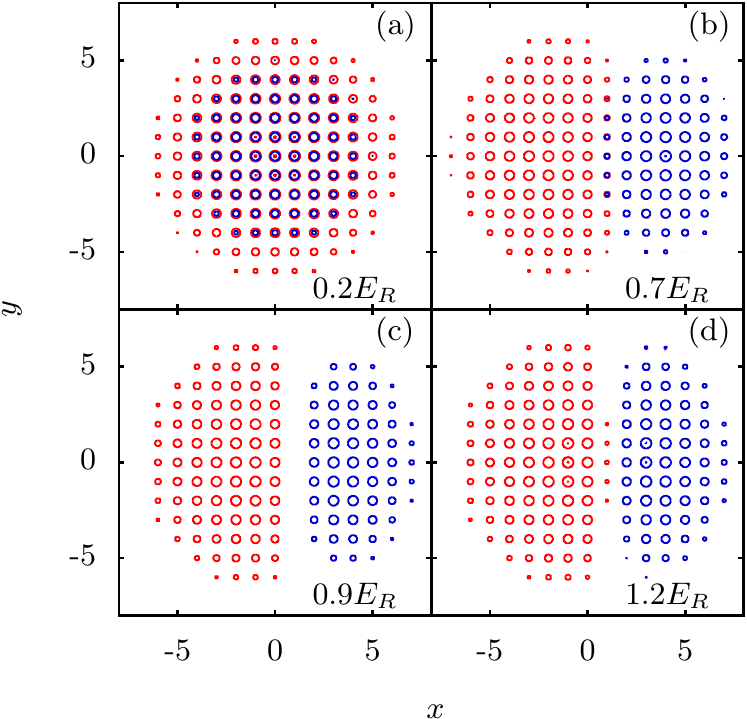}}
  \caption{The geometry of the condensate density profiles and its transition
           from miscible to the immiscible domain in $^{133}$Cs -$^{87}$Rb
           TBEC. Here species labeled $1(2)$ is shown as red (blue) contours.}
  \label{den_cs}
\end{figure}
\begin{figure}[ht]
 {\includegraphics[width=8.5cm] {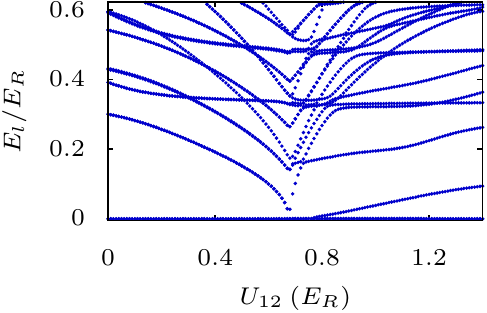}}
  \caption{The evolution of the low-lying modes as a function of the
           interspecies interaction in $^{133}$Cs-$^{87}$Rb TBEC held in
           quasi-2D optical lattices. Here $U_{12}$ is in units of the recoil
           energy $E_R$.}
  \label{mode_cs_en}
\end{figure}

For the $^{133}$Cs - $^{87}$Rb TBEC, as mentioned earlier, we 
vary interspecies interaction $U_{12}$ to induce the miscible to
the immiscible phase transition. 
\begin{figure}[ht]
 {\includegraphics[width=8.5cm] {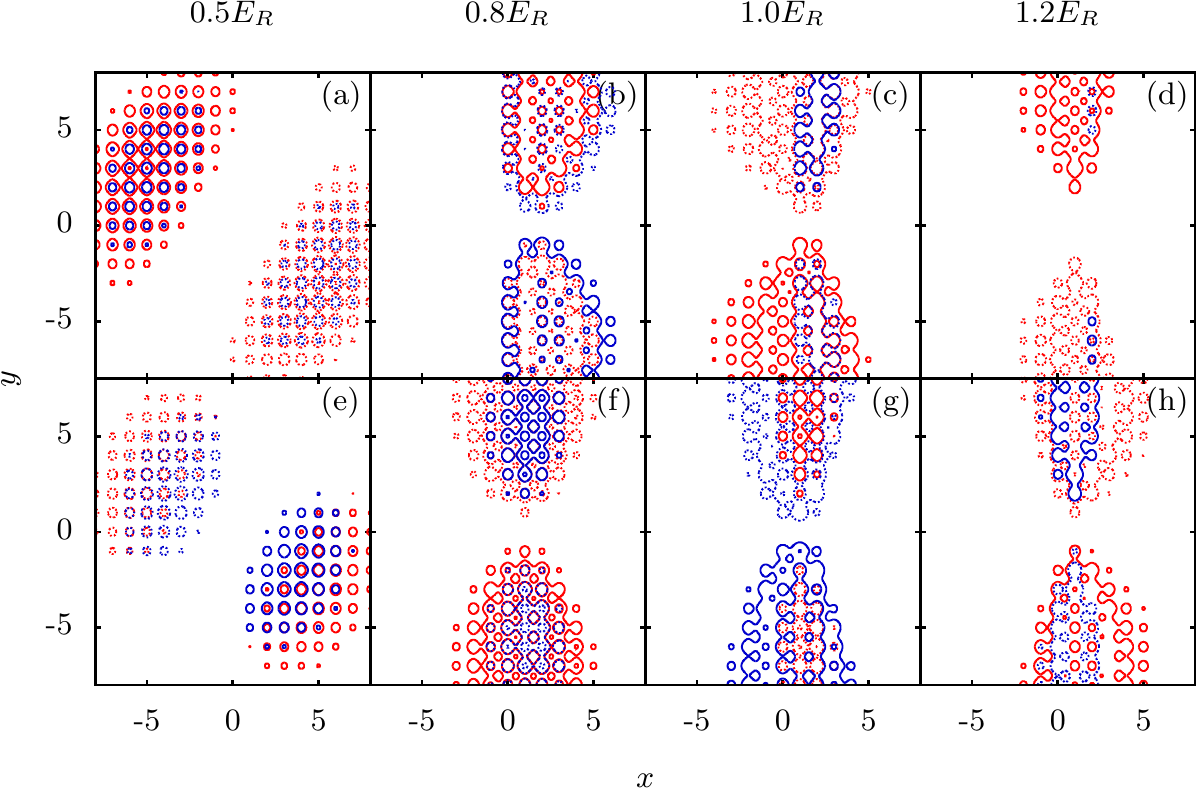}}
  \caption{The evolution of quasiparticle amplitude corresponding to the
           slosh mode of first 
           species (a-d) and second species (e-h) of $^{133}$Cs -$^{87}$Rb
           TBEC as $U_{12}$ is increased from $0.5 E_R$ to $1.2 E_R$. The 
           value of $U_{12}$ is shown at the top of the figures. Here the red 
           contours represent the quasiparticle amplitude ($u_{1}(x,y)$ and 
           $u_{2}(x,y)$), whereas the blue contours represent the quasihole 
           amplitude ($v_{1}(x,y)$ and $v_{2}(x,y)$).  
           }
  \label{mode_fn1_cs}
\end{figure}
\begin{figure}[ht]
 {\includegraphics[width=8.5cm] {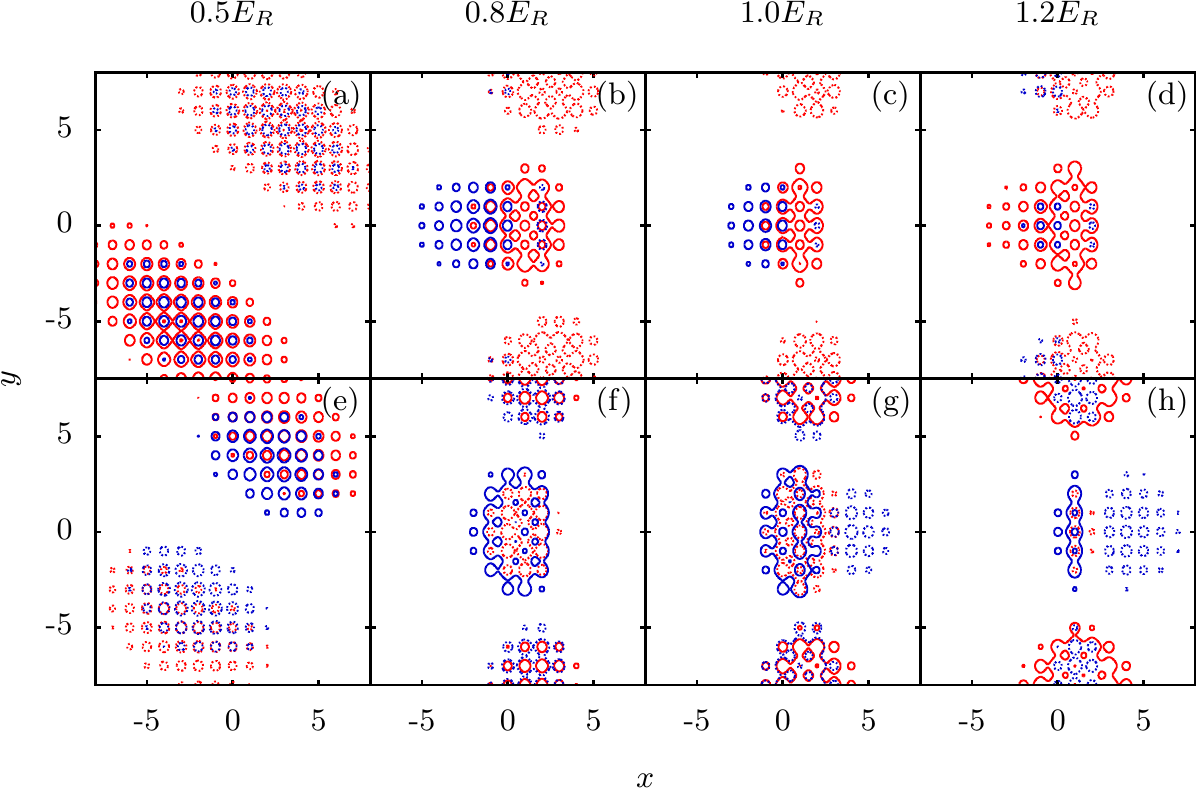}}
   \caption{The evolution of the quasiparticle mode corresponding to the 
            slosh mode, 
            which is degenerate with the mode shown 
            in Fig.~\ref{mode_fn1_cs} in miscible domain. These amplitudes 
            correspond to the first species (a-d) and second species (e-h) of 
            $^{133}$Cs -$^{87}$Rb TBEC as $U_{12}$ is increased from $0.5 E_R$ 
            to $1.2 E_R$. The value of $U_{12}$ is shown at the top of the 
            figures. At a critical value of $U_{12}$, the energy of the mode 
            increases and it gets transformed into an interface mode. 
            Here the red contours represent the quasiparticle amplitude, 
            whereas the blue contours represent the quasihole amplitude. 
            }
  \label{mode_fn2_cs}
\end{figure}
The density profiles, as the miscible-to-immiscible transition occurs, are
shown in Fig.~\ref{den_cs}. The change, except for the curvature at the 
interface, are similar to the case of  $^{87}$Rb - $^{85}$Rb TBEC shown in
Fig.~\ref{den_rb}. The evolution of the mode energies before, during and 
after the transition are shown in Fig.~\ref{mode_cs_en}. Like in the previous
case, $^{87}$Rb - $^{85}$Rb TBEC, the slosh mode is degenerate in the miscible 
domain [shown in Fig.~\ref{mode_fn1_cs}(a,e) and Fig.~\ref{mode_fn2_cs}(a,e)].
It goes soft at the critical value $U^c_{12} = 0.68 E_R$, and the degeneracy 
is lifted. As shown in Fig.~\ref{mode_fn1_cs}(b,c,d,f,g,h) and
Fig.~\ref{mode_fn2_cs}(b,c,d,f,g,h), the evolution of the non-degenerate modes
are qualitatively similar to that of $^{87}$Rb - $^{85}$Rb TBEC.
One key feature in the general trend of the mode evolution is, in the 
miscible domain all the mode energies decrease with increase in $U_{12}$. 
However, as discussed earlier, in $^{87}$Rb - $^{85}$Rb TBEC the energies 
of all the in-phase modes (modes with same phase of $u_1$ and $u_2$) remain 
steady. At phase separation, the mode energies reach minimal values and 
then, increase with increasing $U_{12}$ in the immiscible domain. To gain 
an insight on these trends, we examine the dependence on various parameters 
with a series of computations.

Based on the results, we observe that the form of the interaction,
interspecies or intraspecies, which is tuned to drive the
miscible-to-immiscible transition has an impact on the trends of the mode
evolution. An important observation is, for high $U_{kk}/J_k$ all the modes
decrease in energy, in the miscible domain, when the interspecies interaction
is tuned. However, when the intraspecies interaction is tuned all the 
in-phase modes remain steady. Thus, we attribute the difference in the trends 
to the geometry of the interface at phase separation. When the interspecies 
interaction is tuned, as in $^{133}$Cs - $^{87}$Rb TBEC, the interface at 
phase separation is linear as evident from Fig.~\ref{den_cs}(c). Thus, it can 
align with the nodes of the mode functions, and decrease all the mode 
energies. This is not possible in the other case, tuning intraspecies 
interaction in $^{87}$Rb - $^{85}$Rb, as the interface is curved as shown in 
Fig.~\ref{den_rb}(c).

\begin{figure}[ht]
 {\includegraphics[width=8.5cm] {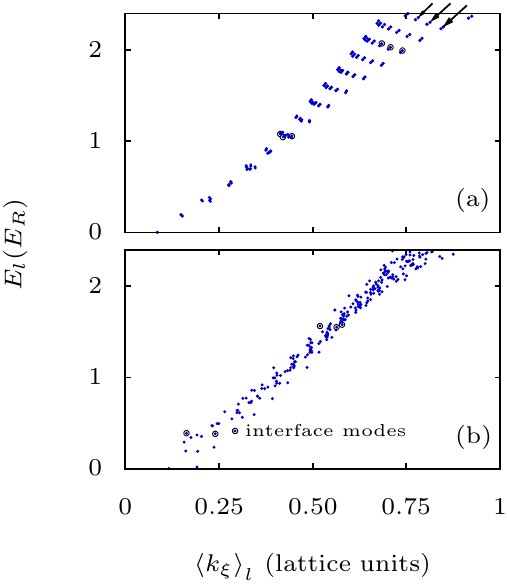}}
  \caption{The discrete BdG quasiparticle dispersion curve in (a) miscible
           and (b) immiscible domain of $^{87}$Rb -$^{85}$Rb TBEC.}
  \label{disp_rb_rb}
\end{figure}


\subsection{Dispersion relations}

 To obtain dispersion curves, based on Eq.~(\ref{wave_vec}), we compute
$\langle k_\xi\rangle_l$ of the $l$th quasiparticle, and plot the mode 
energies. To highlight trends in the dispersion curves dependent on angular
momentum, we choose parameters different from what we have considered
so far. Further more, we restrict ourselves to the case of 
$^{87}$Rb - $^{85}$Rb TBEC, where the trends in dispersion curves are
more prominent due to weaker inter-atomic interactions, and small mass 
difference. In particular, we consider a system of $^{87}$Rb - $^{85}$Rb 
TBEC with DNLSE parameters $J_{1} = J_{2} = 0.66 E_R$, and 
$U_{11} = U_{22} = 0.01 E_R$. For the interspecies on-site interactions 
$U_{12}$, to explore the dispersion relations in miscible and immiscible 
domains we set it to $0.003 E_R$ and $0.08 E_R$, respectively. All the other 
parameters are retained with the same values as mentioned earlier. One 
important point to be emphasized is, unlike the parameters in the mode 
evolutions studies, the current choice of DNLSE parameters correspond to 
two different sets of $N_1$ and $N_2$.

\subsubsection{Miscible domain}

\begin{figure}[ht]
 {\includegraphics[width=8.5cm] {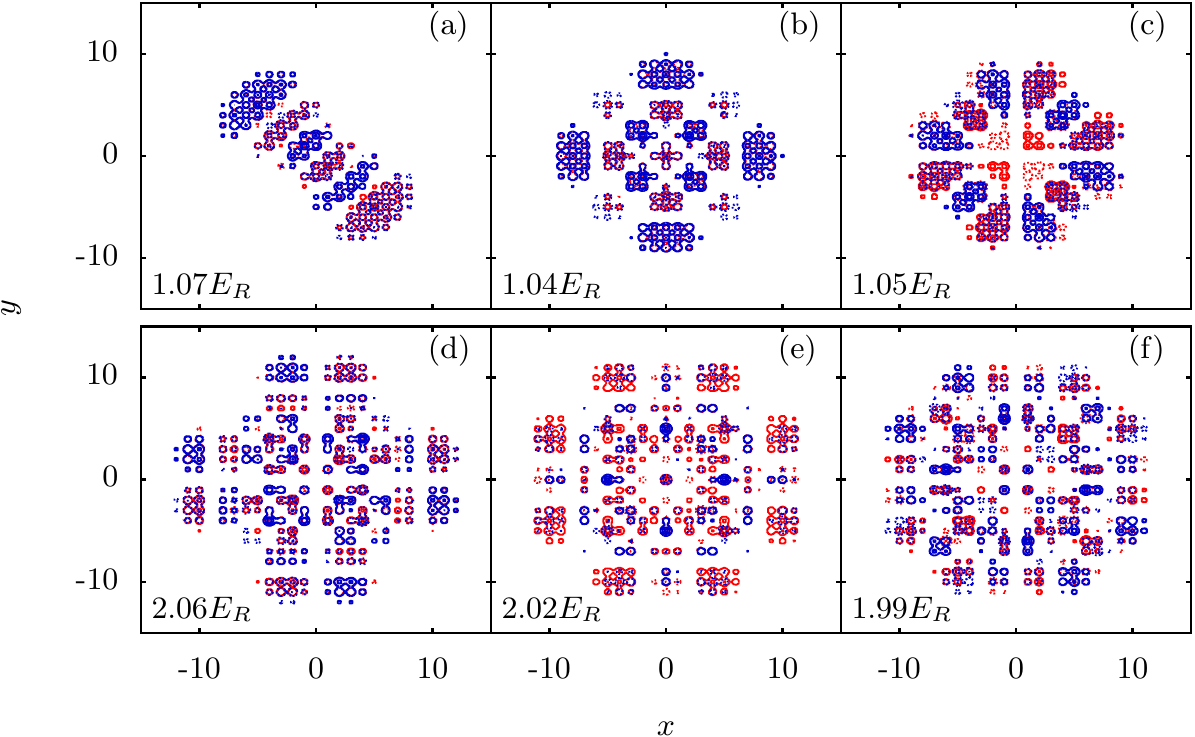}}
  \caption{Shown here are the quasiparticle amplitudes in the 
           miscible domain of a TBEC.
           (a-c) quasiparticle amplitudes with excitation energy 
           ($\approx 1E_R$) and (d-f) quasiparticle amplitudes with
           excitation energy 
           ($\approx 2E_R$). These quasiparticles are indicated 
           in dispersion plot [Fig.~\ref{disp_rb_rb}(a)] by black circles. The 
           excitation energies corresponding to each quasiparticle is written 
           in the lower left corner of each plot in units of the recoil
           energy. Here excitations corresponding to species 1 (2) are
           shown with red (blue) contours.}
  \label{mode_fn_mis}
\end{figure}

 The ground state of the system has rotational symmetry in this domain. 
Hence, the azimuthal quantum number ($m$) is a good quantum number, 
and finite interspecies interaction mixes modes with same $m$ arising 
from each of the two species. This is reflected in the branch like
structures in the dispersion curve as shown in Fig.~\ref{disp_rb_rb}(a).
To understand the physics behind the structure of the dispersion curves, 
we examine the structure of the quasiparticle modes. For this, let us
focus on modes which lie on three branches, marked by arrows,  
in Fig.~\ref{disp_rb_rb}(a). Each of the modes can be identified based on 
the value of $m$. As example, three of the low-energy ($\approx 1E_R$) and 
another three from higher energies ($\approx 2E_R$) are shown in 
Fig.~\ref{mode_fn_mis}. 

The energies of the first three quasiparticle modes in the figure, 
Fig.~\ref{mode_fn_mis}(a-c), are out-of-phase type, and the values 
of $m$ are 1, 4 and 6. Among these modes, the first two modes have
$\langle k_\xi\rangle_l\approx 0.42$, and are phonon-like as these lie on 
the linear part of the dispersion curve. However, the mode in 
Fig. ~\ref{mode_fn_mis}(c) with $\langle k_\xi\rangle_l\approx 0.44$ and 
$m=6$ is a surface mode, which is evident from the structure of the
mode function. The same observation is confirmed from the exponential decay
in the numerical values of $u$ towards the center. These three modes show
that within the same energy range ($\approx 1E_R$), phononlike and
surface excitation co-exists. One discernible trend is, the modes
with higher $m$ and $\langle k_\xi\rangle_l$ have extremas located farther 
from the center of the trap, and turn into surface modes. The quasiparticle 
amplitudes with higher excitation energies ($\approx 2 E_R$), shown 
in Fig.~\ref{mode_fn_mis}(d,e,f), have intricate structures. This is as
expected arising from the larger mode mixing due to higher density of 
states and non-zero $U_{12}$.

\begin{figure}[ht]
 {\includegraphics[width=8.5cm] {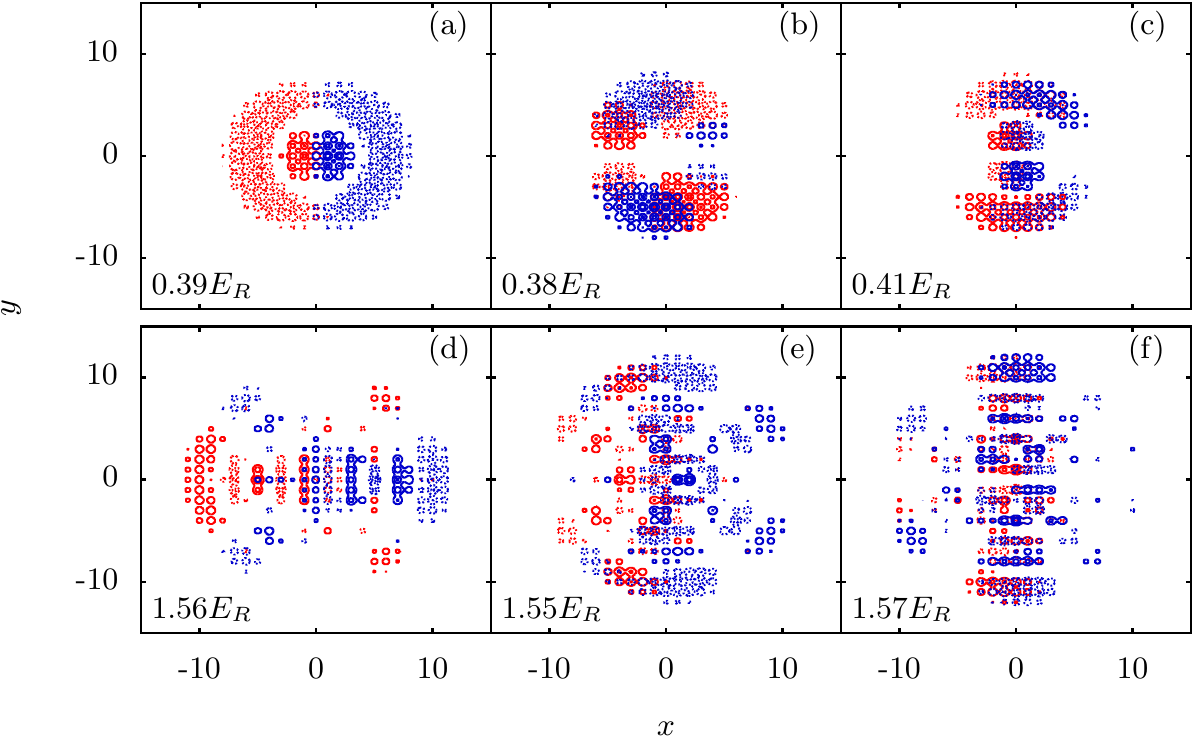}}
  \caption{Shown here are the quasiparticle amplitudes in the
           immiscible domain of a TBEC.
           (a-c) quasiparticle amplitudes with excitation energy
           ($\approx 0.4E_R$) and (d-f) quasiparticle amplitudes with
           excitation energy
           ($\approx 1.5E_R$). These quasiparticles are indicated
           in dispersion plot [Fig.~\ref{disp_rb_rb}(b)] by black circles. The
           excitation energies corresponding to each quasiparticle is written
           in the lower left corner of each plot in units of the recoil
           energy. Here excitations corresponding to species 1 (2) are
           shown with red (blue) contours.}
  \label{mode_fn_immis}
\end{figure}

\subsubsection{Immiscible domain}

 For the immiscible domain, the dispersion curve is as shown in 
Fig.~\ref{disp_rb_rb}(b), and there are no discernible trends.
The reason is, in this domain the condensate density profile does not have
rotational symmetry, and hence, there are mixing between quasiparticle
modes with different $m$-values. To examine the structure of the mode
functions we consider three each with energies $\approx 0.4E_R $ and
$\approx 1.55 E_R$, these are shown in  
Fig.~\ref{mode_fn_immis}(a-c), and Fig.~\ref{mode_fn_immis}(d-f), 
respectively. Consider the modes with energies $0.39E_R$ and $0.38E_R$ as
shown in Fig.~\ref{mode_fn_immis}(a), and (b), the flow patterns in these
are equivalent to the breathing and slosh modes in single species
condensates, respectively. There is, however, one important difference:
the density flow involves both the species, and have different velocity
fields. The mode with energy $0.41E_R$, shown in Fig.~\ref{mode_fn_immis}(c),
is out-of-phase in nature and has a different configuration compared to 
the two previous ones. That is, the mode functions are prominent around the 
interface region, and are negligible in the region where the condensate 
densities are maximal. In continuum case, modes with similar structure
(interface mode) has been reported in recent works 
~\cite{ticknor_13,ticknor_14}. The mode with higher energies have enhanced
mode mixing due to higher density of states, which is evident from the
structure of the modes with $\approx 1.55 E_R$ shown in 
Fig.~\ref{mode_fn_immis}(d-f). Hence, it is non-trivial to classify the
modes like in the case of modes with energies $\approx0.4E_R$. In terms of
the geometrical structures, the modes in Fig.~\ref{mode_fn_immis}(d), (e),
and (f) have extrema coincident with the condensates, interlaced distribution,
and localized in the interface region, respectively. Thus, within a range of
excitation energies, there exists modes with diverse characters. 


\section{Conclusions}
\label{conc}
  Our studies show that the introduction of an optical lattice potential
modifies the geometry of condensate density distribution of TBECs at phase 
separation. The sandwich or shell structured density profiles are no longer
energetically favourable, and the side-by-side geometry emerges as the only
stable ground state density profile. This arises from the higher interface 
energy due to the local density enhancements at lattice sites. The other
important observation is, as the TBEC is tuned from miscible to immiscible
phase, the evolution of the quasiparticle spectra can be grouped into two.
The first group have quasiparticles which exhibit a decrease in the mode
energies as we approach phase-separation, and reach minimal values at the
critical interaction strength. However, the mode energies increase
after crossing into the  domain of phase-separation. The second group, on
the other hand, remains steady as the interaction strength is tuned across 
the critical value. Furthermore, we have examined the dispersion curves 
for miscible and immiscible domains of TBEC. The curves, in the miscible 
domain,  show discernible trends associated with the azimuthal quantum 
number of the quasiparticle. However, in the immiscible domain, there
are no discernible trends associated with azimuthal quantum number. This is
due to the rotational symmetry breaking of the condensate density profiles, 
and the resulting mixing of modes  with different azimuthal quantum numbers.


\begin{acknowledgments}
 We thank Arko Roy, S. Gautam, S. Bandyopadhyay and R. Bai for useful 
discussions. The results presented in the paper are based on the computations 
using Vikram-100, the 100TFLOP HPC Cluster at Physical Research Laboratory, 
Ahmedabad, India.
\end{acknowledgments}

\bibliography{tbec_2d_opl}{}
\bibliographystyle{apsrev4-1}
\end{document}